\documentclass[twocolumn,aps,prd,showpacs,preprintnumbers,amsmath,amssymb,nofootinbib]{revtex4}
\usepackage{graphicx}
\usepackage{dcolumn}

\newcommand{\AF}{{\it Asymfast}}
\newcommand\order[1] { ${{\cal O}\! \left( #1 \right)}$ }
\begin{document}

\title{Asymfast, \\a method for convolving maps with asymmetric main beams}

\author{M.~Tristram}
\email{tristram@lpsc.in2p3.fr}
\affiliation{Laboratoire de Physique Subatomique et de Cosmologie, 53 Avenue des Martyrs, 38026
  Grenoble Cedex, France}
\author{J-Ch.~Hamilton}
%\affiliation{Laboratoire de Physique Subatomique et de Cosmologie, 53 Avenue des Martyrs, 38026
%  Grenoble Cedex, France}
\affiliation{Laboratoire de Physique Nucl\'eaire et de Hautes
  Energies, 4 place Jussieu,  75252 Paris Cedex 05, France}
\author{J.~F.~Mac\'\i as-P\'erez}
\affiliation{Laboratoire de Physique Subatomique et de Cosmologie, 53 Avenue des Martyrs, 38026
  Grenoble Cedex, France}
\author{C.~Renault}
\affiliation{Laboratoire de Physique Subatomique et de Cosmologie, 53 Avenue des Martyrs, 38026
  Grenoble Cedex, France}

\date{\today}
   
\begin{abstract}
  We describe a fast and accurate method to perform the convolution of
  a sky map with a general asymmetric main beam along any given
  scanning strategy. The method is based on the decomposition of the
  beam as a sum of circular functions, here Gaussians. It can be
  easily implemented and is much faster than pixel-by-pixel
  convolution. In addition, {\AF} can be used to estimate the
  effective circularized beam transfer functions of CMB instruments
  with non-symmetric main beam. This is shown using realistic
  simulations and by comparison to analytical approximations which are
  available for Gaussian elliptical beams. Finally, the application
  of this technique to Archeops data is also described. Although
  developped within the framework of Cosmic Microwave Background
  observations, our method can be applied to other areas of
  astrophysics.
\end{abstract}


\maketitle

%________________________________________________________________

\section{Introduction}

With the increasing in accuracy and angular scale coverage of the
recent Cosmic Microwave Background (CMB) experiments, a major
objective is to include beam uncertainties when estimating
cosmological parameters \cite{bridle}, and in particular the asymmetry
of the beam \cite{arnau}. As seen in {\sc Cosmosomas}
\cite{cosmosomas}, {\sc Boomerang} \cite{boom}, {\sc Maxima}
\cite{maxima}, {\sc Archeops} \cite{archeops}, {\sc WMAP} \cite{wmap}
and anticipated for {\sc Planck}, the systematics errors are dominated
at the smaller scales (higher multipoles, $\ell$) by the uncertainties
on the reconstruction of and deconvolution from the beam pattern. So
the beam must be considered more realistically, in particular by
rejecting the assumptions of Gaussianity and/or symmetry. It leads to a
better modelization of the beam to convolve with and a better
estimation of its representation in the harmonic space, its transfer
function $B_{\ell}^{eff}$.

The use of simulated sky maps takes an important part in the
analysis of astrophysical data sets to study possible systematic
effects and noise contributions, and also, to compare the data to
theoretical predictions or to observations from other instrumental
setups. This is one of the most challenging problems in data
analysis~: obtaining a simple and accurate model for the instrumental
response. In many cases, Monte-Carlo approaches are favoured as they
are in general simpler than the analytic ones. However, they need a
great number of simulations which often requires too much time of
execution for the available computing facilities. A large amount of
this time is spent in the convolution of the simulated data by the
instrumental response or beam pattern. For an asymmetric beam, the
convolved map at a given pointing direction on the sky would depend
both on the relative orientation of the beam on the sky and on the
shape of the beam pattern. This makes brute-force convolution particularly
painful and slow ({\it e.g.} \cite{buriganaLFI}).

Therefore, either we work in the spherical harmonic space using a
general and accurate convolution algorithm (see \cite{wandelt} for a
fast implementation), either we model the beam pattern by a series of
easy-to-deal-with functions and compute for each a fast convolution in
the harmonic space. 

The work of~\cite{wandelt} presents how to convolve {\it exactly} two
band limited but otherwise arbitrary functions on the sphere - which
can be the 4-$\pi$ beam pattern and the sky descriptions. At each
point on the sphere, one computes a ring of different convolution
results corresponding to all relative orientations about this
direction. To allow subsequent interpolation at arbitrary locations it
is sufficient to discretize each Euler angle describing the position
on the sphere into \order{L} points where L measures the larger of the
inverse of the smallest length scale of the sky or beam. The method
rapidity depends on the scanning strategy : in \order{L^3} for
constant-latitude scans and in \order{L^4} for other strategy - even if
factorization may be found in some cases. This method is used for
Planck simulations in case of elliptical Gaussian beams for the
polarized channels \cite{whansen} and is one of the most promising
techniques to decrease side-lobe effects on the WMAP results
\cite{barnes}. While it is efficient for a certain class of
observational strategies \cite{whansen}, it may be difficult to
implement in the more general case of non-constant latitude scanning
strategies \cite{challinor}.

For the second method based on the modelling the beam pattern, several
solutions have been proposed either circularizing the beam
(\cite{page} or \cite{wu}) or assuming an elliptical Gaussian beam
(\cite{souradeep} or \cite{pablo}). Here we propose an alternative
method, {\AF}, which easily can account for any main beam shape. In
general the main beam maps can be obtained from point-sources like the
planets Jupiter and Saturn (see \cite{chiang} for a general view,
\cite{beamplanck}, \cite{beamlfi} for HFI, LFI beams and \cite{page}
for WMAP beams). The beam maps are then fitted with the appropriate
model from which the beam transfer function can be computed. Moreover
the beam modelization can be used as input for deconvolution methods
like \cite{burigana} or optimal map-making iterative methods as
MapCUMBA \cite{dore}.

By contrast to \cite{wandelt}, {\AF} deals only with main beams which
are decomposed onto a sum of 2D symmetric Gaussians and does not take
into account far-side lobes. In the following, we will simply call
Gaussian a 2D symmetric Gaussian function. The sky is then easily
convolved by each Gaussian and the resulting convolved sub-maps are
combined into a single map which is equivalent to the sky convolved by
the asymmetric original beam. This method can be used for any
observational strategy and is easy to implement. As the goal of {\AF}
is only to deal with main beams, we shall not directly compare with the
\cite{wandelt} method in this paper.

The use of 4-$\pi$
beam is beyond the scope of {\AF} and so {\AF} and the \cite{wandelt}
method are not directly compared in this paper.

We describe our approach in Section~\ref{method} and the simulations we
used to check the accuracy and performance of the method in
Section~\ref{simu}. Section~\ref{fit} describes the symmetric
expansion of the beam. In Section~\ref{precision}, we compare the
accuracy of our method with respect to the elliptical Gaussian and the
brute-force approach. Section~\ref{time} discusses the time-computing
efficiency of the different convolution method considered. Finally, in
Section~\ref{bell}, we describe a method to estimate the effective
beam transfer function. This technique has been successfully used
for the determination of the Archeops main beam in section~\ref{archeopstot}.

\section{Method \label{method}}
{\AF} approaches any asymmetric beam by a linear combination of
Gaussian functions centered at different locations within the original
beam pattern.

The convolution is performed separetly for each Gaussian and then the
convolved maps are combined into a single map. The map convolution
with the Gaussian beams is computed in the spherical harmonic space.
This allows us to perform the convolution in a particularly low time
consuming way. By contrast, the brute-force convolution by an asymmetric
beam needs to be performed in the real space for each of the time
samples so that the relative orientation of the beam on the sky is
properly taken into account for each pointing direction.

The {\AF} method can be described in five main steps~:
\begin{enumerate}
\item The beam is decomposed into a weighted sum of $N$ Gaussians. The
  number of Gaussians is chosen by minimazing residuals according to
  user-defined precision (see Section~\ref{fit}).
  
\item The initial map is oversampled\footnote{we consider oversampled
sub-maps to reduce the influence of the pixelisation.}  by a factor of
2 and convolved with each of the Gaussian functions. The sky map is
decomposed into $a_{\ell m}$ coefficients in the harmonic space. Then,
$N$ sub-maps are computed, each of them smoothed with the
corresponding Gaussian sub-beam, by multiplying the $a_{\ell m}$
coefficients of the original sky map by the current approximation of
the transfer functions of the Gaussian beams $B_\ell^N=\exp \left[
-\ell\left(\ell+1\right)\sigma_N^2 \right]$.
  
\item The sub-maps are deprojected into timelines using the scanning
  strategy of the corresponding sub-beam position.
  
\item The $N$ timelines are stacked into a single one weighted by
  their sub-beam amplitude.
  
\item The timeline obtained this way is projected onto the sky with
  the scanning strategy corresponding to the center of the beam
  pattern and we obtain the convolution of the original sky map with
  the fitted beam along the scan.
\end{enumerate}

The HEALPix \cite{healpix} package is used to store maps ({\em ring}
description), to compute the decomposition of the sky map in spherical
harmonics ({\em anafast}) and to reconstruct maps from the $a_{\ell
  m}$ coefficients ({\em synfast}).

\section{Beams and scanning strategy simulations \label{simu}}

The method was checked using realistic simulations of a sky
observation performed by an instrument with asymmetric beam pattern
and complex scanning strategy. To keep the time consumption
reasonnable within our computing capabilities without degrading the
quality of the results, we have chosen beams with FWHM between 40 and
60~arcmin sampled by steps of 4~arcmin (square beam map of
60$\times$60~pixels corresponding to 4$\times$4~deg).

We consider two sets of simulations~: a first one
corresponding to a quasi-circular beam to which we have added 1~\%
(simulation 1a) and 10~\% (simulation 1b) of noise, and a second one
corresponding to an irregular beam to which we have also added noise
in the same way (simulations 2a and 2b).
These two sets are obtained from a sum of a random number of
elliptical Gaussians with random positions within 0.4~degree around
the center and random FWHMs and amplitudes. Then, the beams are
smoothed with a 4~arcmin width Gaussian and white noise is added.

The scanning strategy is assumed to be a set of consecutive meridians
with sampling of 3~arcmin over each meridian and a lag of 3~arcmin
between two meridians. This strategy is roughly similar to the
Planck observing plan. It allows an efficient check of the method as,
close to the equator, all beams are parallel and the effect of the
orientation is maximum. However, close to the pole, the high number of
beams in different direction per pixel makes the effective beam more
circular. In order to save time, we keep only half an hemisphere of
the sky, so we use about 6.5 million data samples. The number of hits
per pixel is highly variable from equator to pole, ranging from 1 to
about 8000 near the pole.

For high-resolution instruments, like {\sc Planck}, we should consider
maps with resolution of the order of 1~arcmin. However, this would
require too much computing time for brute-force convolution. Instead,
in this paper, we consider maps of the sky with pixel size of
$\approx$24~arcmin which are stored in the HEALPix format ($N_{side}=256$). We
have checked that the results obtained with {\AF} considering lower
resolution maps with larger beams can be generalized to higher
resolution maps with narrower beams.

\section{Symmetric Gaussian expansion of the beam \label{fit}}

We model the original beam pattern using $N$ Gaussians, $i \in
\{1,N\}$, of the form
\begin{equation}
g(x_i,y_i,\sigma_i;x,y) = \frac{1}{2 \pi \sigma_i^2} \exp \left[
    \frac{(x-x_i)^2 + (y-y_i)^2}{2 \sigma_i^2} \right ]
\end{equation}
The beam is fitted with a weighted sum of the $N$ Gaussians~:
\begin{equation}
b(x,y) = \sum_{i=1}^N A_i g(x_i,y_i,\sigma_i;x,y)
\end{equation}
so we have $4N$ free parameters corresponding to width ($\sigma_i$,
\mbox{FWHM = $\sqrt{8 ln2} \sigma_i$}), amplitude ($A_i$) and
center position of each Gaussian ($x_i, y_i$).

The optimal value of $N$ depends on the required precision. We perform
the fit using 1 to 10 Gaussians (which is typically enough to attain
residuals of the order of the noise level) and compute the quadratic
deviation to the original beam pattern
\begin{equation}
S(N) = \frac{1}{n_p} \sum_{p=1}^{n_p} \left[b^{fit}_N(p) - b(p)\right]^2.
\end{equation}
\noindent where $n_p$ represents the total number of pixels,
$b^{fit}_N$ corresponds to the fitted beam with $N$ Gaussians and $b$
to the original beam pattern.

\begin{figure*}
\begin{center}
\resizebox{!}{22cm}{
  \includegraphics[angle=90]{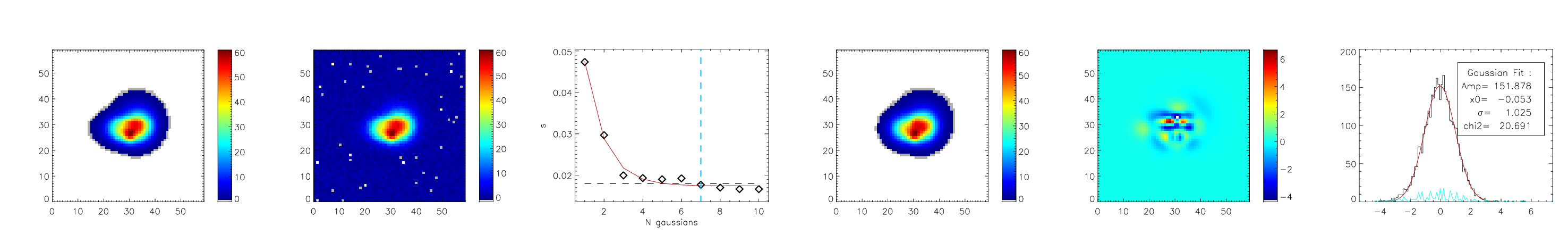}
  \includegraphics[angle=90]{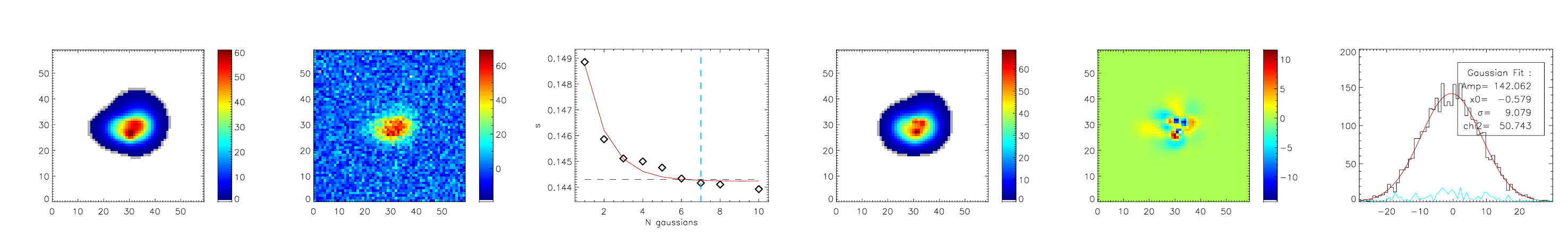}
\hspace{2cm}
  \includegraphics[angle=90]{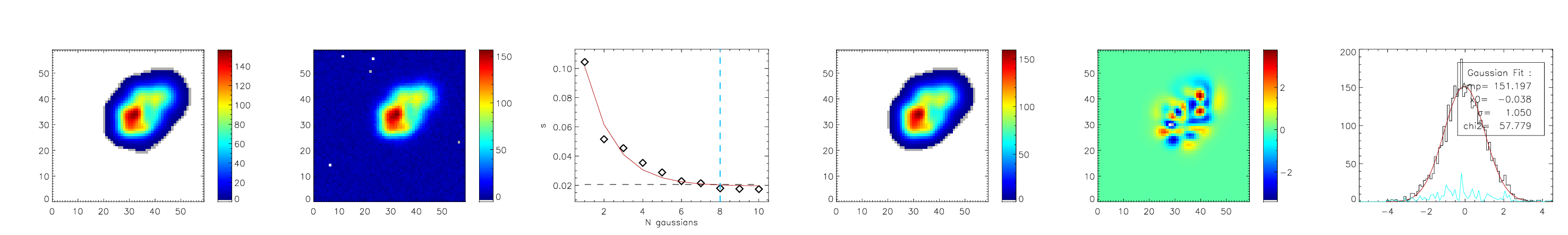}
  \includegraphics[angle=90]{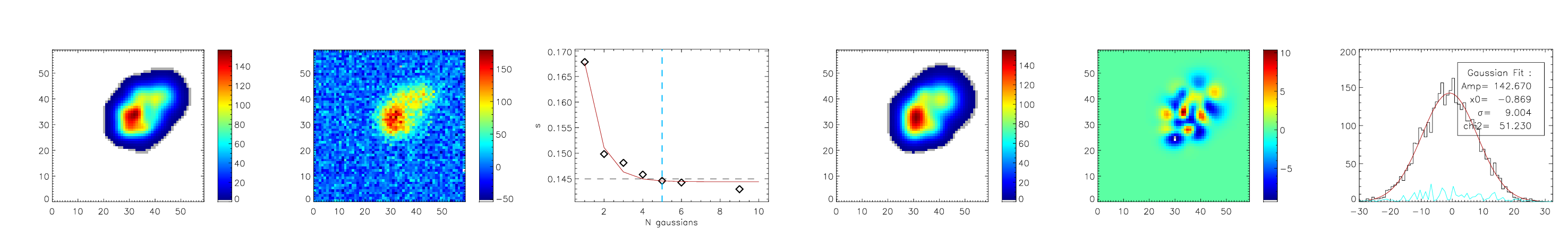}
}
\caption{\label{fig_simu}Main results for the two sets of simulations
  described in Section~\ref{fit} (first two rows simulations 1a and 1b,
  last two rows simulations 2a and 2b). From left to right for each
  row~: the initial simulated beam pattern, {\it id} + noise, the
  quadratic deviation $S$ as a function of the number of Gaussians,
  the best-fit model for the beam pattern, the residual map after
  substraction of the input noise in percents and the histogram of the residual
  map including input noise. The dot blue lines show the computed
  number of Gaussian necessary to reproduce the initial simulated
  beam. The histogram of the residual map is fitted with a Gaussian
  (in red), the difference is shown in blue.}
\end{center}
\end{figure*}

We consider now the two sets of simulations (simulations 1a, 1b, 2a
and 2b) described in Section~\ref{simu}. The fit is performed using a
least square fit in the non-linear case following the algorithm
described in \cite{brandt}. Figure~\ref{fig_simu} presents the main
results for these two sets of simulations~: the first two rows for
simulations 1a and 1b, and the last two rows for simulations 2a and
2b. In each case, we have plotted, from left to right~:
\begin{itemize}
\item the initial simulated beam pattern
\item the initial simulated beam pattern + noise
\item the quadratic deviation $S$ as a function of the number of
  Gaussians $N$ (the dotted line represents the number of Gaussians
  chosen for the final beam pattern model)
\item the best-fit model for the beam pattern
\item the residual map after substraction of the input noise in percents
\item the histogram of the residual map including input noise, fitted
  to a Gaussian (in red) with residuals to the Gaussian (in blue)
\end{itemize}

The distribution of $S(N)$ is modelled by a decreasing exponential,
$exp(-\tau N)$, plus a constant $k$. The number of Gaussians $N$ is
chosen such that $N$ is the smaller value verifying
\begin{equation}
S(N)-k < \frac{t}{\tau}
\end{equation}
\noindent where $t$ is a threshold defined by the user for the
required precision.

Figure~\ref{fig_simu} shows that a small number of Gaussians is enough
for a good fit (typically less than 10~Gaussians give less than 2~\%
residuals). In some cases, the algorithm does not converge and
therefore no data point is plotted on the figure. Lack of convergence
may appear for some combination of number of Gaussians. This is
generally solved by using a smaller or larger number of Gaussians in
the fit.

The efficiency of the fit is illustrated by the distribution of the
residuals which are very close to the white noise distribution
centered at zero with the dispersion corresponding to the level of the
input noise. Note also that the reconstructed beams are marginally
sensitive to the noise level and that the apparent decomposition in
three different peaks (case 1b) leads to residuals with amplitude only
of the level of the noise.

Once the Gausssians parameters have been estimated, map simulations
using any pointing strategy can be done quickly and precisely.

\section{Accuracy of the Asymfast method \label{precision}}

In this section, we compare the accuracy of {\AF} to that of other
common approaches including the modelling of the beam pattern either
by a single circular Gaussian or by an elliptical Gaussian.
We use as a reference the brute-force convolution with the
true beam pattern at each pointing direction.

For this purpose, we have convolved, using the different methods
described before, a map containing 26 point sources with the same
amplitudes uniformly distributed over half an hemisphere. The level of
accuracy for each of the former methods of convolution is estimated by
computing the quadratic deviation of the convolved map obtained for
that particular method with respect to the convolved map obtained by
brute-force convolution.

\begin{table}[t]
  \renewcommand{\arraystretch}{1.4}
  \begin{center}
  \begin{tabular}{c|ccc}
    \hline
    \hline
beam       &  1 circular  & 1 elliptical & {\it asymfast} \\ 
simulation &   Gaussian   &   Gaussian   &  (N Gaussians)   \\ \hline
     1a &   4.444 &   1.187 &   0.734 (7)\\
     1b &   4.420 &   1.281 &   0.936 (7)\\
     2a &  10.513 &   4.095 &   0.826 (8)\\
     2b &  10.464 &   4.154 &   1.455 (5)\\
\hline
  \end{tabular}
  \caption{
    Accuracy of the three methods discussed in the text (see
    Section~\ref{precision}) for each set of simulations.
    \label{results_rms}}
  \end{center}
\end{table}

Table~\ref{results_rms} represents the quadratic deviation for each of
the methods as a percentage of the maximum of the original
point-sources map. This quantity can be interpreted as the percentage
of spurious noise introduced by the convolution method. In the case of
quasi-circular beams (simulations 1a and 1b), {\AF} is 1.5 (resp.~6)
more accurate than the elliptical (resp. circular) approximation. For
more realistic irregular beams (simulations 2a and 2b), {\AF} is about 4 (resp. 10) times more
accurate than the elliptical (resp. circular) approach. Moreover {\AF}
depends neither on the shape of the beam pattern nor on the noise
level.

\begin{figure}[t]
\resizebox{\hsize}{!}{
  \hspace{-2cm}
  \includegraphics[]{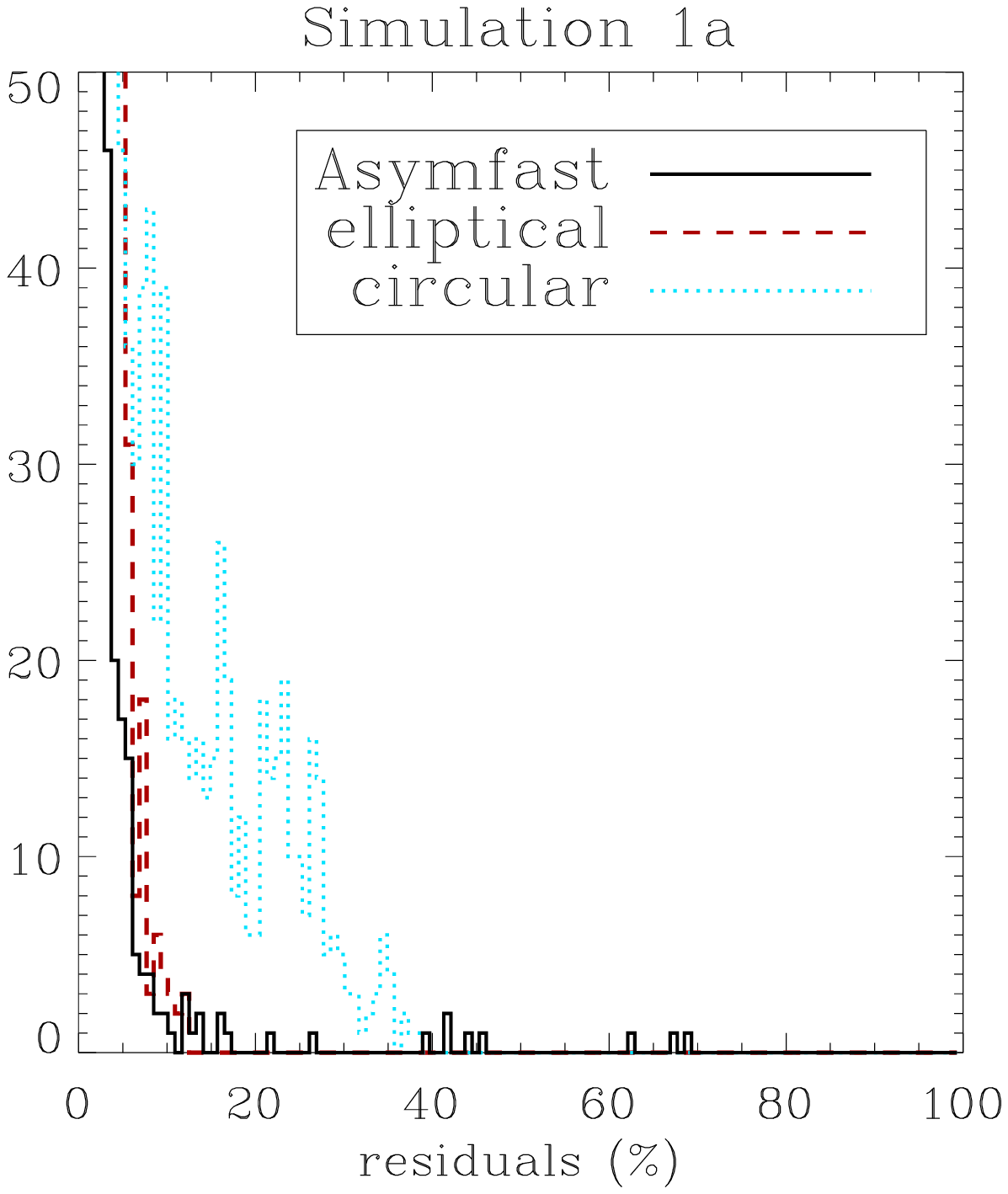}
  \includegraphics[]{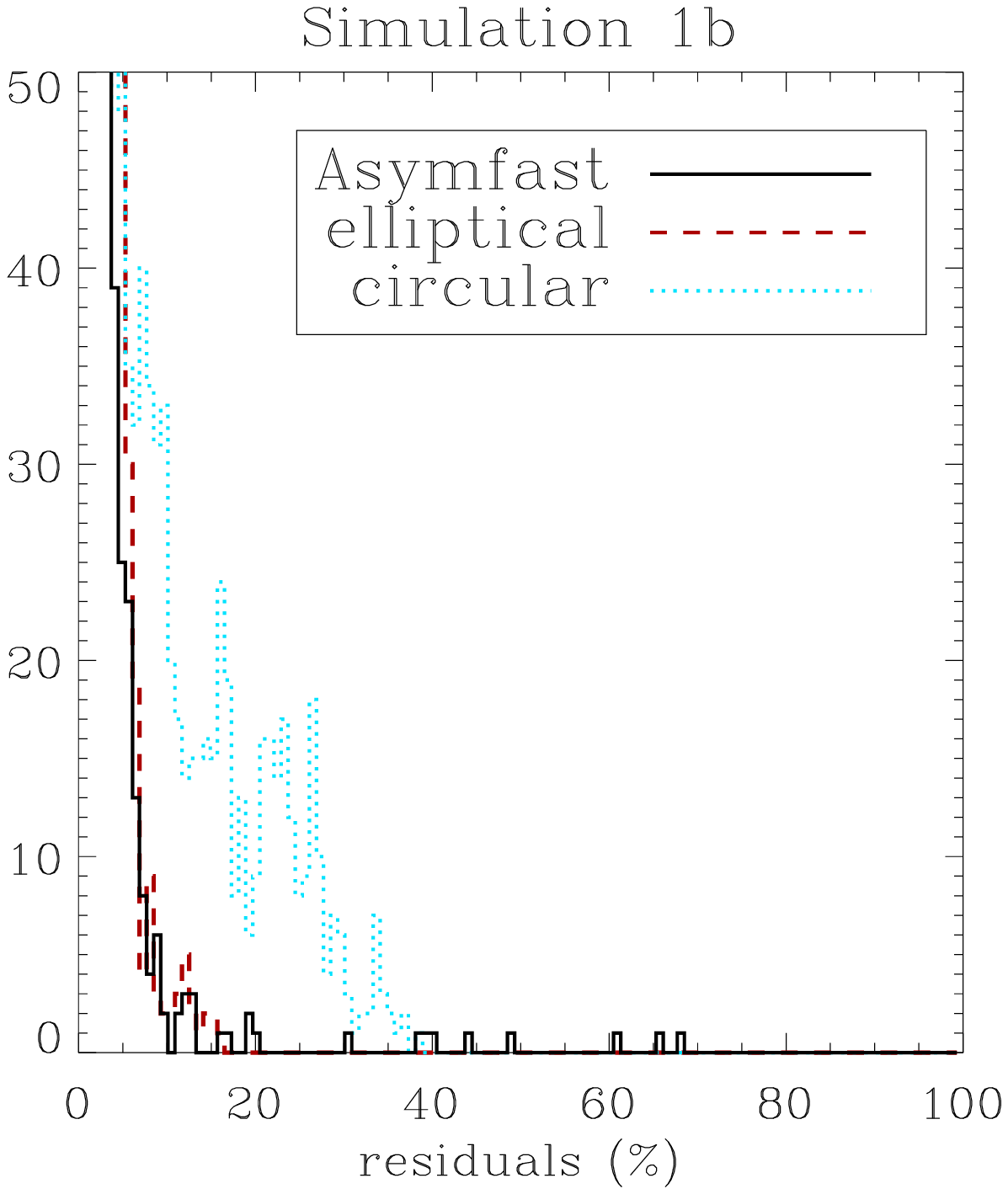}}
\resizebox{\hsize}{!}{
  \hspace{-2cm}
  \includegraphics[]{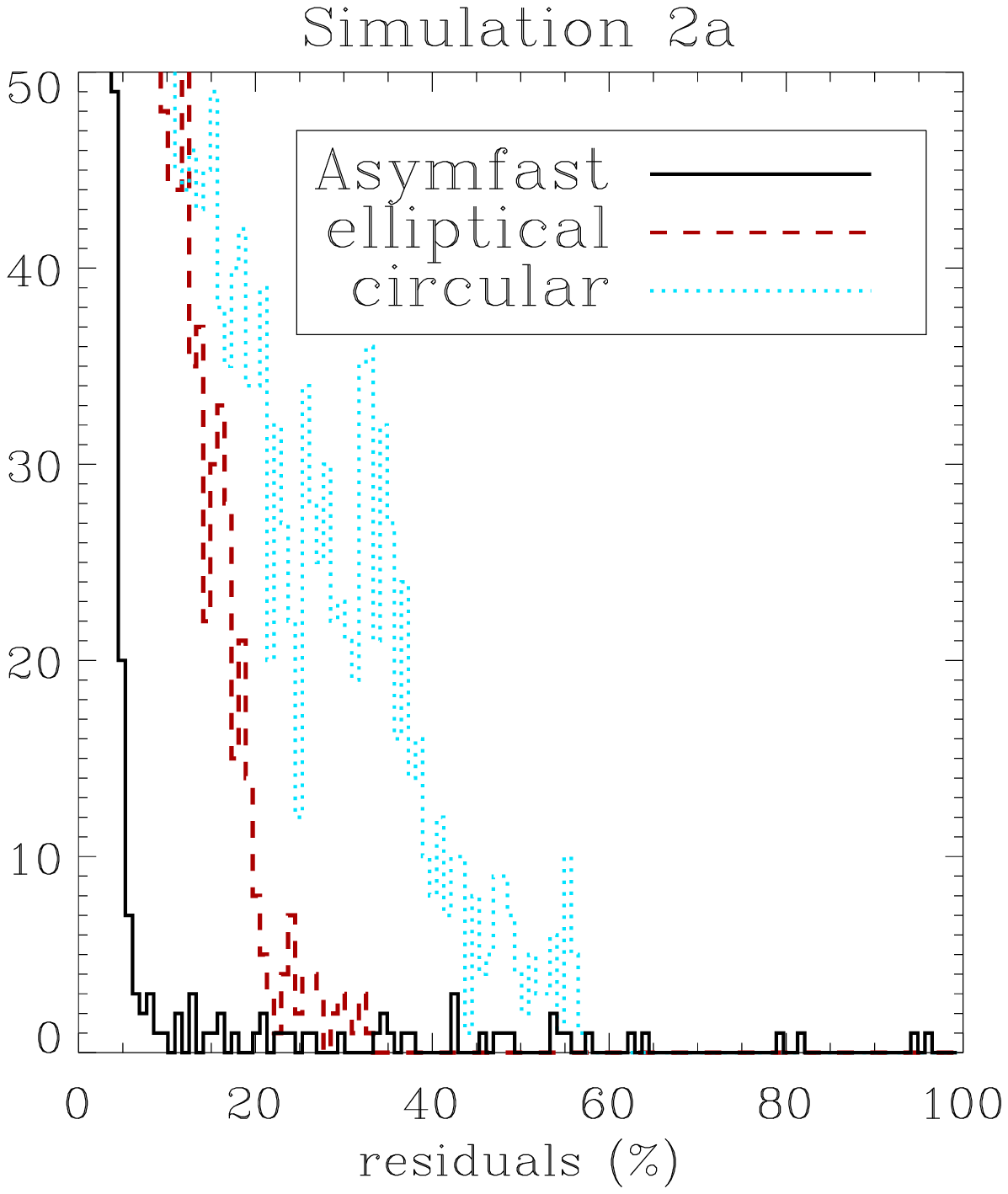}
  \includegraphics[]{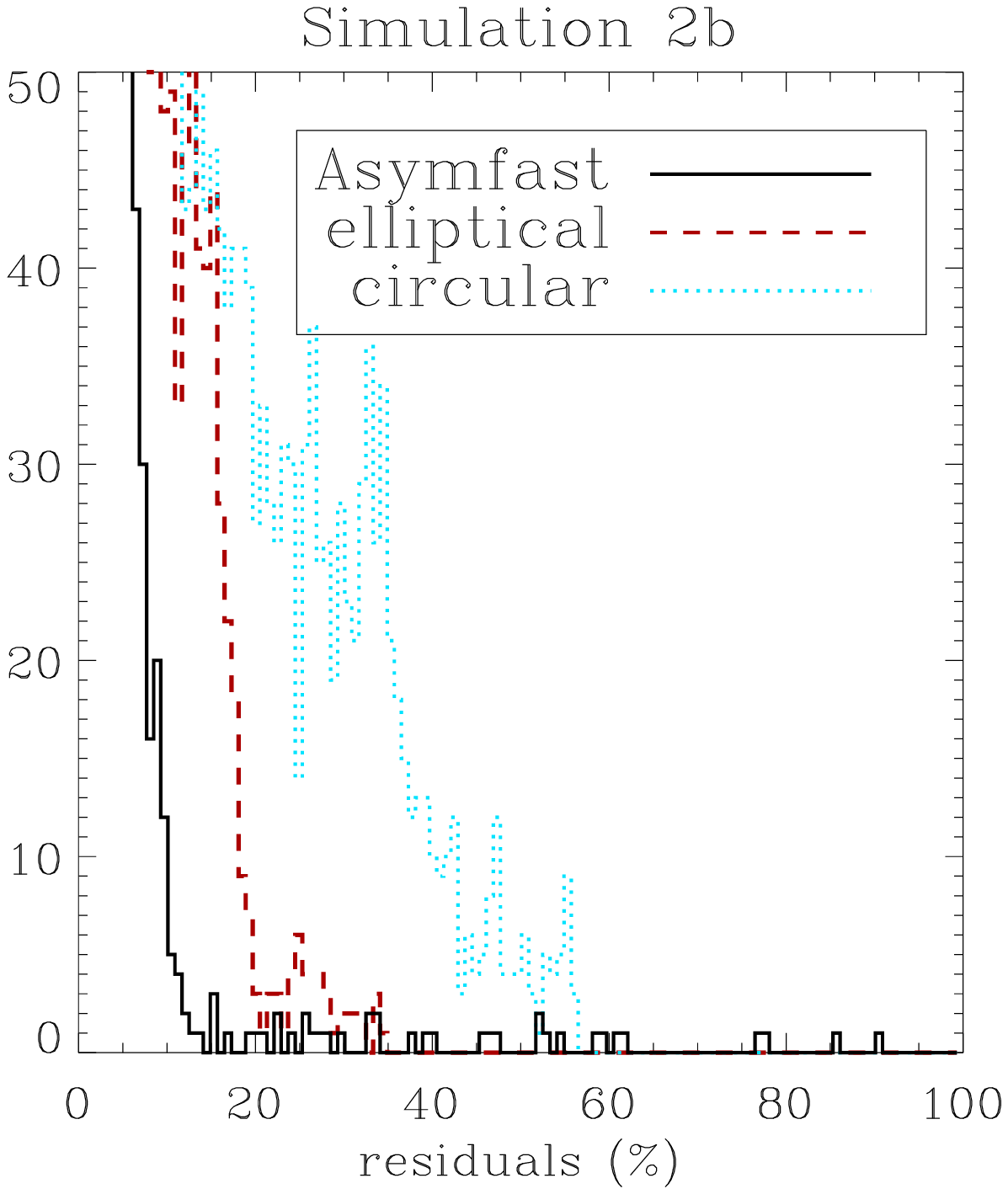}}
  \caption{Histogram of the percentage of the quadratic deviation residual
    maps for the {\AF} (solid line), elliptical (dashed line) and
    circular (dotted line) approximations. Top panel~: simulations 1a,
    1b. Bottom panel~: simulations 2a, 2b.\label{fig_histo}}
\end{figure}

Figure~\ref{fig_histo} shows, for the two sets of simulations
considered (simulations 1a, 1b on the top panel and simulations 2a, 2b
and the bottom panel), the histogram of the percentage of the
quadratic deviation residual maps for the {\AF} (solid line),
elliptical (dashed line) and circular (dotted line) approximations.
The figure confirms the results shown in Table~\ref{results_rms},
indicating that {\AF} is a very good approximation to both
quasi-circular and irregular beams (residuals smaller than 5\%). By
contrast, the elliptical approximation can only be safely used in the
case of quasi-circular beams (residuals about~20\% for simulations 2).
The circular approximation, as expected, is very poor in any case
(residuals of the order 40\%).

In conclusion, the {\AF} results are very close to the standard
solution while the circular and elliptical Gaussian approximations are
very poor for realistic beam patterns. In addition, the standard
approach, which works in real-space, necessarily uses a beam pattern
with fixed resolution. By contrast, {\AF} works in the spherical
harmonic space where the resolution is only limited by the
pixelisation of the sky map. Moreover, {\AF} can obtain full
resolution in frequency with no mixing. Obtaining the same frequency
resolution with brute-force convolution requires a full-sky beam pattern
which increases considerably the computing time.

\section{Time-computing efficiency of Asymfast {\it vs} brute-force 
convolution \label{time}}

As discussed in the following section, the determination in the
spherical harmonic space of the effective circular transfer function of
an asymmetric beam pattern, $B_{\ell}^{eff}$, requires a large number
(of the order of 1000) of accurate Monte-Carlo simulations of fake
data appropriately convolved by the instrument beam. Therefore, any
algorithm used for this purpose need to be much faster than the
brute-force convolution procedure to keep the computing time reasonable.

To quantify the time-computing efficiency of {\AF}, we consider
$n_{simu}=1000$ Monte-Carlo simulations of a convolved full-sky map
for a single {\sc Planck}-like detector. We use a squared map of
$n_{beam}=3600$ pixels to describe the beam pattern and simulated
timelines for 12 months of {\sc Planck} observations which correspond
to $n_{scan} = 5\ 10^9$~samples. With {\AF}, we model the beam pattern
using $N=10$ Gaussians. For each simulation, the total number of
pixels\footnote{We have considered the HEALPix convention
  \cite{healpix}} on the full-sky map is $n_{pix} = 12 * N_{side}^2$
for three values of $N_{side}$~: 256, 512 and 1024 considered.

\begin{table}[t]
\begin{center}
\renewcommand{\arraystretch}{1.4}
\begin{tabular}{c|ccc}
\hline
\hline
 Nside  & HEALPix & {\it Asymfast} & brute-force       \\
  &   1 Gaussian  & 10 Gaussians  & convolution \\ \hline
  256 & 0.127 & 7.40  & 171  \\
  512 & 1.02 & 29.1  & 1340  \\
 1024 & 8.13 & 203  & 10700  \\
\hline
\end{tabular}
\caption{ Computing-time, in terms of $10^6$ arbitrary CPU-units and
including the input/output time access, for circular beam convolution
in the harmonic space using the HEALPix package, and for asymmetric
beam convolution using {\AF} and brute-force convolution algorithms. We
have considered a set of 1000 simulations as described in
Section~\ref{time}.
  \label{time_results}}
\end{center}
\end{table}

For these simulations, Table~\ref{time_results} shows the
computing-time, in terms of arbitrary CPU-units, for circular beam
convolution in the harmonic space using the HEALPix package, and for
asymmetric beam convolution using {\AF} or brute-force convolution
algorithms.

We observe that {\AF} is more than 50~times faster than the brute-force
convolution for the high resolutions ($N_{side} = 512$ ou $1024$).
This is because the convolution in the spherical harmonic space used
by {\AF} is much faster than in real space. In addition, {\AF} is
particularly efficient for Monte-Carlo purposes because the beam
modelization and the computing of the pointing directions
corresponding to each of the $N$ sub-beam are performed only once.

Furthermore, for high resolutions, the {\AF} computing-time is just a
factor of 2 larger than the computing-time needed for the convolution
of $N$ circular Gaussians in spherical harmonic space. The extra
computing time with respect to the convolution of $N$ circular
Gaussians comes from the projection and the deprojection operations
performed by {\AF}.

Finally, the difference in computing-time between {\AF} and the brute-force
convolution also increases with the beam map resolution as the
second method depends linearly on beam map number of pixels while
{\AF} is very marginally sensitive to it, only at the beam fit step.

\section{Application to CMB analysis~: estimation of the beam transfer
 function $B_{\ell}$ \label{bell}}

\begin{figure*}[t]
\begin{center}
  \resizebox{18cm}{7cm}{
    \includegraphics{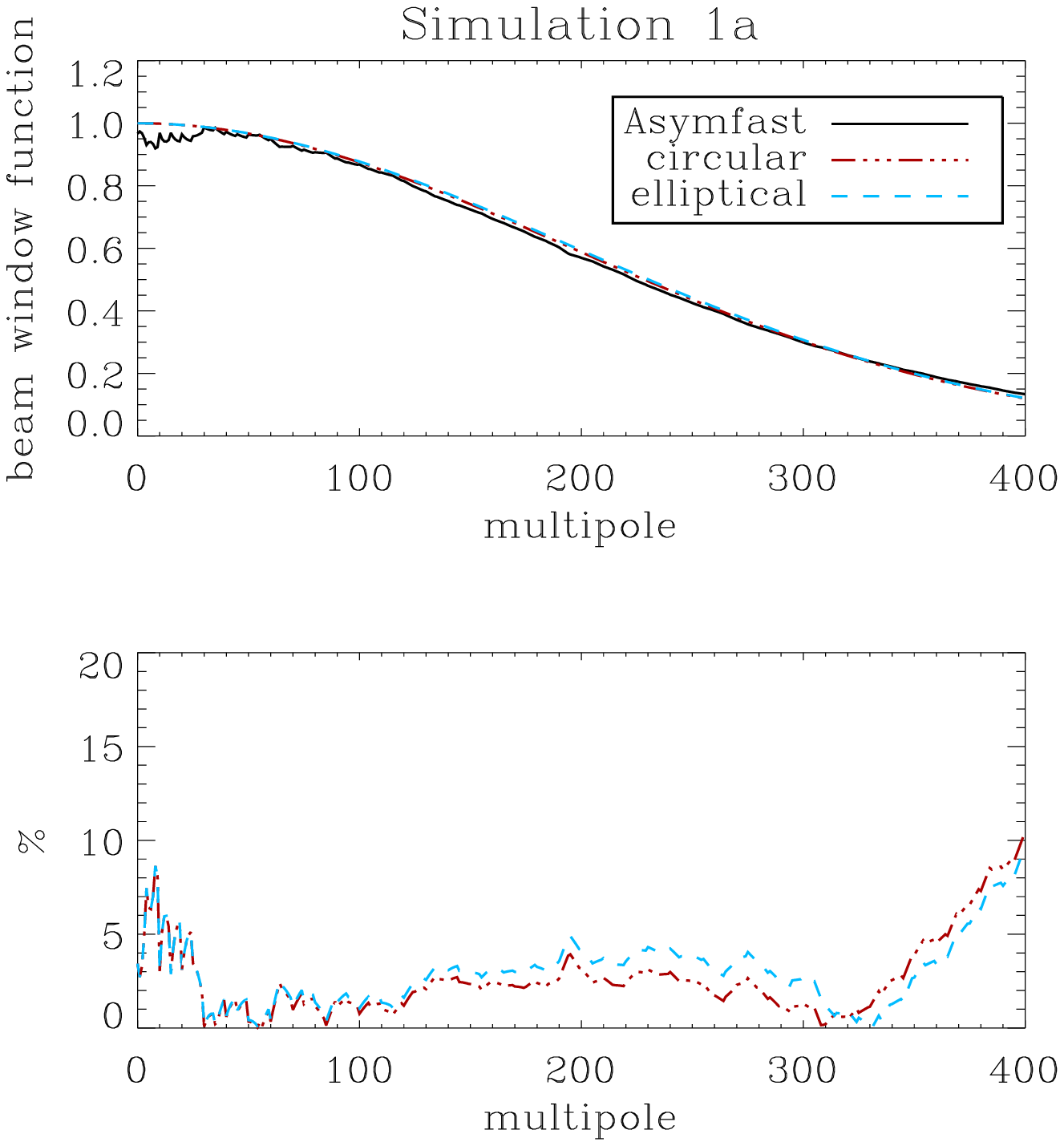}
    \hspace{0.5cm}
    \includegraphics{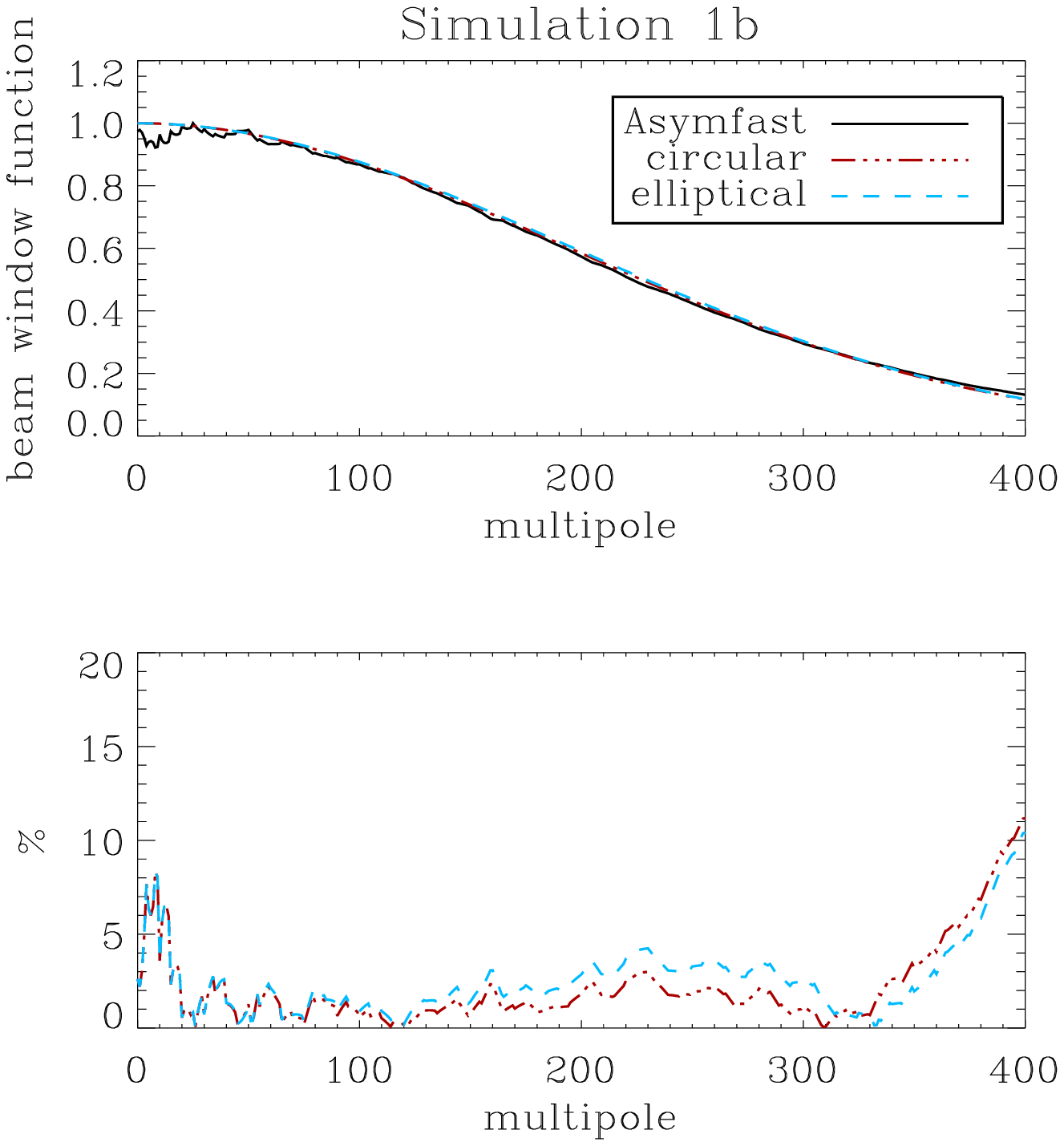}}
\end{center}
\vspace{0.5cm}
\end{figure*}
\begin{figure*}[t]
\begin{center}
  \resizebox{18cm}{7cm}{ \includegraphics{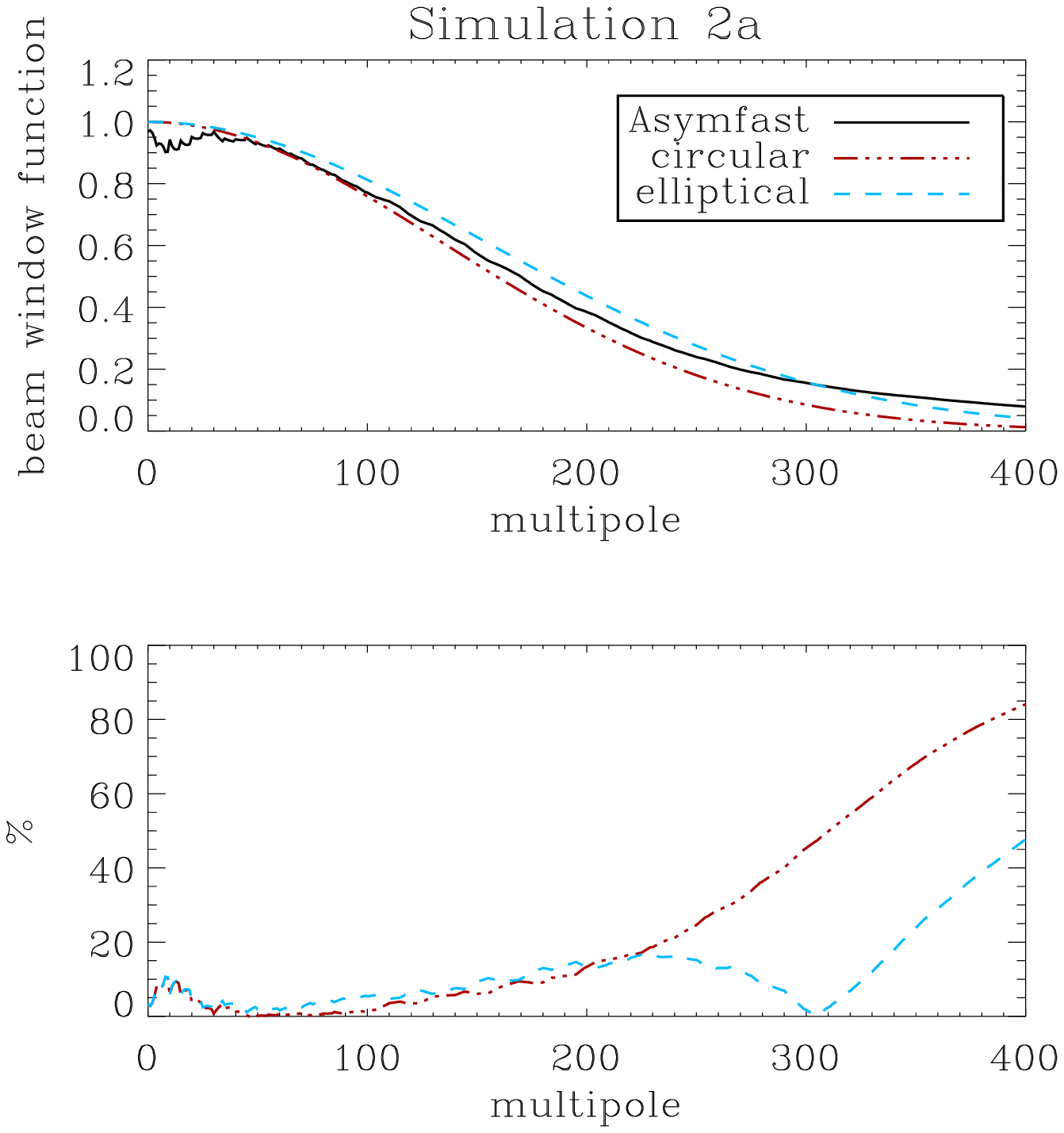}
    \hspace{0.5cm} \includegraphics{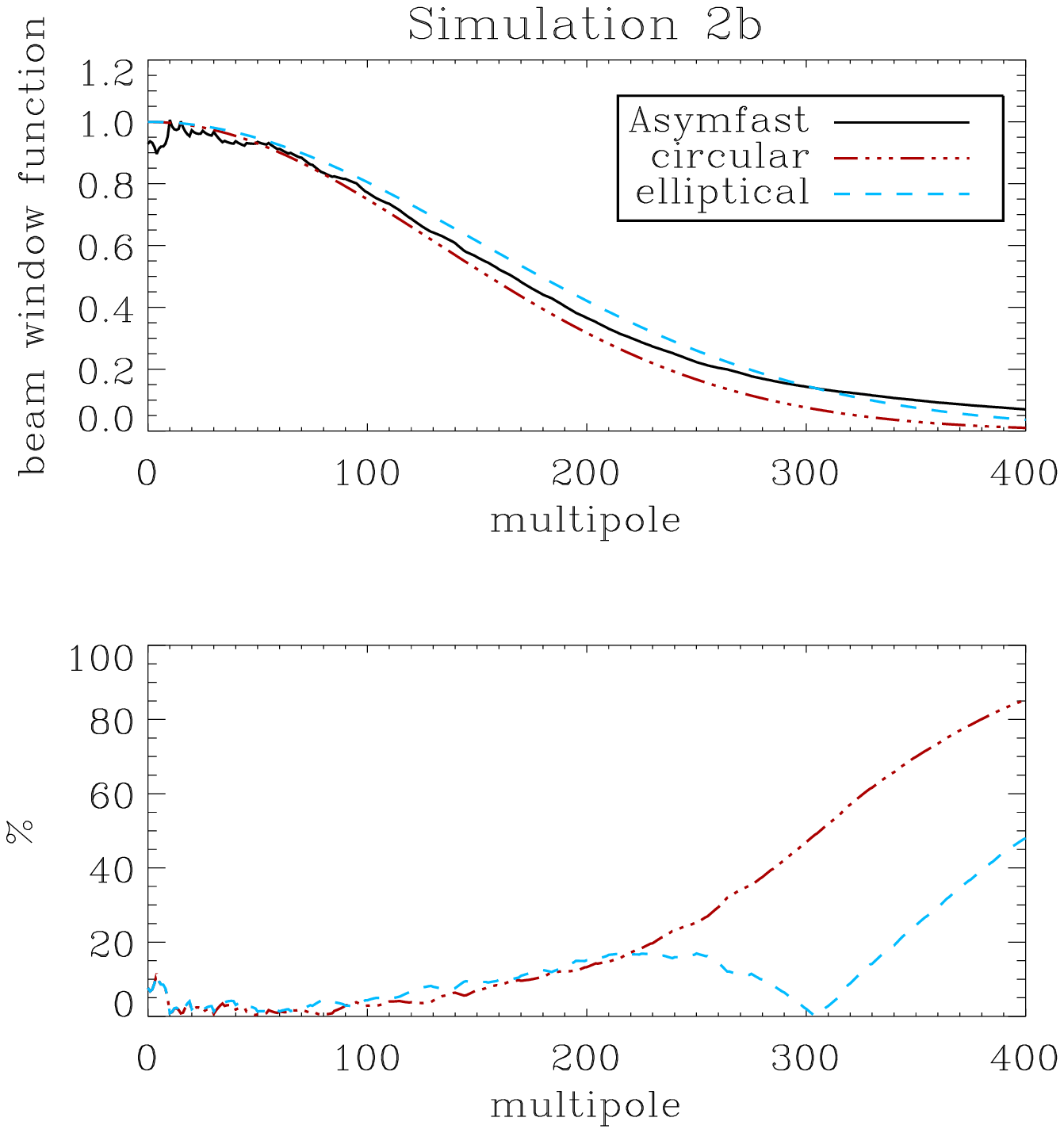}} \caption{Beam
    transfer functions and variations with respect to {\AF} for the
    two sets of simulations (simulations 1a, 1b, 2a and 2b). The
    Monte-Carlo estimation using {\AF} described in Section~\ref{bell}
    (solid line) is compared to the analytic transfer function of a
    circular approximation of the beam (semi-dashed line) and to the
    elliptical approximation to order 4 from~\cite{pablo} (dashed
    line). The top panel represents the beam transfer functions
    whereas the bottom panel shows the differences to the {\AF}
    estimation for the other two approximations. \label{error_bell}}
\end{center}
\end{figure*}

In this section, we describe how {\AF} can be used to estimate this
effective circular transfer function for realistic asymmetric beams.
A beam is described in the spherical harmonic space by a set of
coefficients $b_{\ell m}$ for each scanning orientation at each
pointing position. The Gaussianity of the primordial fluctuations of
the CMB is a key assumption of modern cosmology, motivated by simple
models of inflation. So the angular power spectrum of the CMB should
be a circular quantity and it is common to just consider an effective
circular beam transfer function, $B_{\ell}^{eff}$. The
``circularization''of the beam may be obtained by several ways :
assuming a Gaussian beam (the easiest way, leading to a simple
analytic description), computing the optimal circularly symmetric
equivalent beam (\cite{wu}, applied to Maxima) or fitting the radial
beam profile by a sum of Hermite polynomials (\cite{page}, applied to
WMAP). The analytic expression of the beam transfer function of
elliptical Gaussian beams has also been studied in details by
\cite{pablo} and by \cite{souradeep} who applied it to Python~V.

With {\AF}, the ``circularization''of the beam depends on the scanning
strategy as it is computed on the map after convolution by the
asymmetric main beam. So it is comparable to or more precise than
other methods, depending on how the scans intersect each other.

For a circular Gaussian beam (angles $\theta$ and $\phi$ corresponding
to the beam direction),
\begin{equation}
  b(\theta , \phi) = \frac{1}{2 \pi \sigma^2} \exp
  \left[ - \frac{\theta^2}{2 \sigma^2} \right],
\label{b_theta}
\end{equation}

\noindent the beam transfer function in case of a pixel scale much
smaller than the beam scale is approximated by \cite{white} 
\begin{eqnarray}
B_{\ell} & = & \exp \left[ - \frac{1}{2} \ell (\ell +1) \sigma^2 \right].
\label{bl_gauss}
\end{eqnarray}
Note that in this case, the beam transfer function does not depend on
the orientation of the beam on the sky and is given by a simple
analytical expression.

For a moderate elliptical Gaussian beam pattern, we can compute an
analytic approach \cite{pablo} to the beam transfer function by
introducing a small perturbation to Eq.~\ref{b_theta} so that
\begin{equation}
  b^{ell}(\theta , \phi) = \frac{1}{2 \pi \sigma^2} \exp
  \left[ - \frac{\theta^2}{2 \sigma^2} f(\phi) \right]
\end{equation}
where $f(\phi)$ describes the deviation from circularity. 

For general irregular asymmetric beams, the previous approximation is
not well adapted and the orientation of the beam on the sky has to be
taken into account. For this purpose, we estimate an effective beam
transfer function, $B_{\ell}^{eff}$, from Monte-Carlo simulations
which use the {\AF} method for convolution as follows~:
\begin{enumerate}
\item We convolve CMB simulated maps by the beam pattern
  using {\AF}.
\item For each simulation, we estimate a
  $B_{\ell}^{eff}$ inverting the equation \cite{master}
  $$\tilde{C}_{\ell} = M_{\ell \ell^{\prime}}
  {B_{\ell^{\prime}}^{eff}}^2 C_{\ell^{\prime}}$$
  where $M_{\ell
    \ell^{\prime}}$ is the coupling kernel matrix that takes into
  account the non-uniform coverage of the sky map, $\tilde{C}_{\ell}$
  is the pseudo-power spectrum computed on the convolved map and
  $C_{\ell}$ is the input theorical model.
\item We compute the effective transfer function of the beam, by
  averaging the $B_{\ell}^{eff}$ obtained for each of the simulations.
\end{enumerate}

We have tested this method on simulations of pure circular Gaussian
beams for which we have obtained a beam transfer function fully
compatible with Eq.~\ref{bl_gauss} as expected.

Figure~\ref{error_bell} shows the estimate of the effective beam
transfer function $B_{\ell}^{eff}$ for the two sets of beam simulations
discussed in the previous sections (simulations 1a, 1b, 2a, 2b) using
a Monte-Carlo of 25 simulations. The results of this Monte-Carlo
(solid line) is compared to the analytic transfer function of a circular
approximation of the beam (semi-dashed line) and to the elliptical
approximation computed from~\cite{pablo} (dashed line). The top
panel represents the beam transfer functions whereas the bottom panel
shows the differences to the {\AF} estimation for both analytical
approximations (circular, semi-dashed line and elliptical, dashed line).
The {\AF} effective beam transfer function shows some irregularities
at very low $\ell$ due to the cosmic variance.

As expected, for the quasi-circular beam (simulations 1a and 1b), the
estimate transfer functions for the three methods are similar within
10~\%. The differences observed between our approach and the
elliptical approximation are mainly due to the fact that the latter
does not take into account the scanning strategy. Due to the complex
but realistic scanning strategy used in our simulations, we expect
that, in general, a given position on the sky will be observed with
different relative beam orientations and therefore the effective beam
will appear more circular.

By contrast, for a more irregular beam pattern (simulations 2a and
2b), the differences between the three methods are much larger (from
10~\% at low $\ell$ up to 70~\% at high $\ell$). The beam transfer
function obtained using the elliptical approximation follows better
the one obtained using {\AF}. We expect that the complex scanning
strategy would make the effective beam more circular. However as the
beam pattern is very asymmetric, the effective beam will contain
complex highly irregular structures which cannot be mimicked by an
oriented elliptical beam. Therefore, the largest differences are found
at high resolution.

Note that the results presented here are also valid for higher
resolution beams. For this paper, we have considered low resolution
beams to be able to directly compare the {\AF} convolution to standard
brute-force convolution.

\section{Application to Archeops\label{archeopstot}}

\begin{figure}[h]
  \resizebox{\hsize}{!}{\includegraphics[]{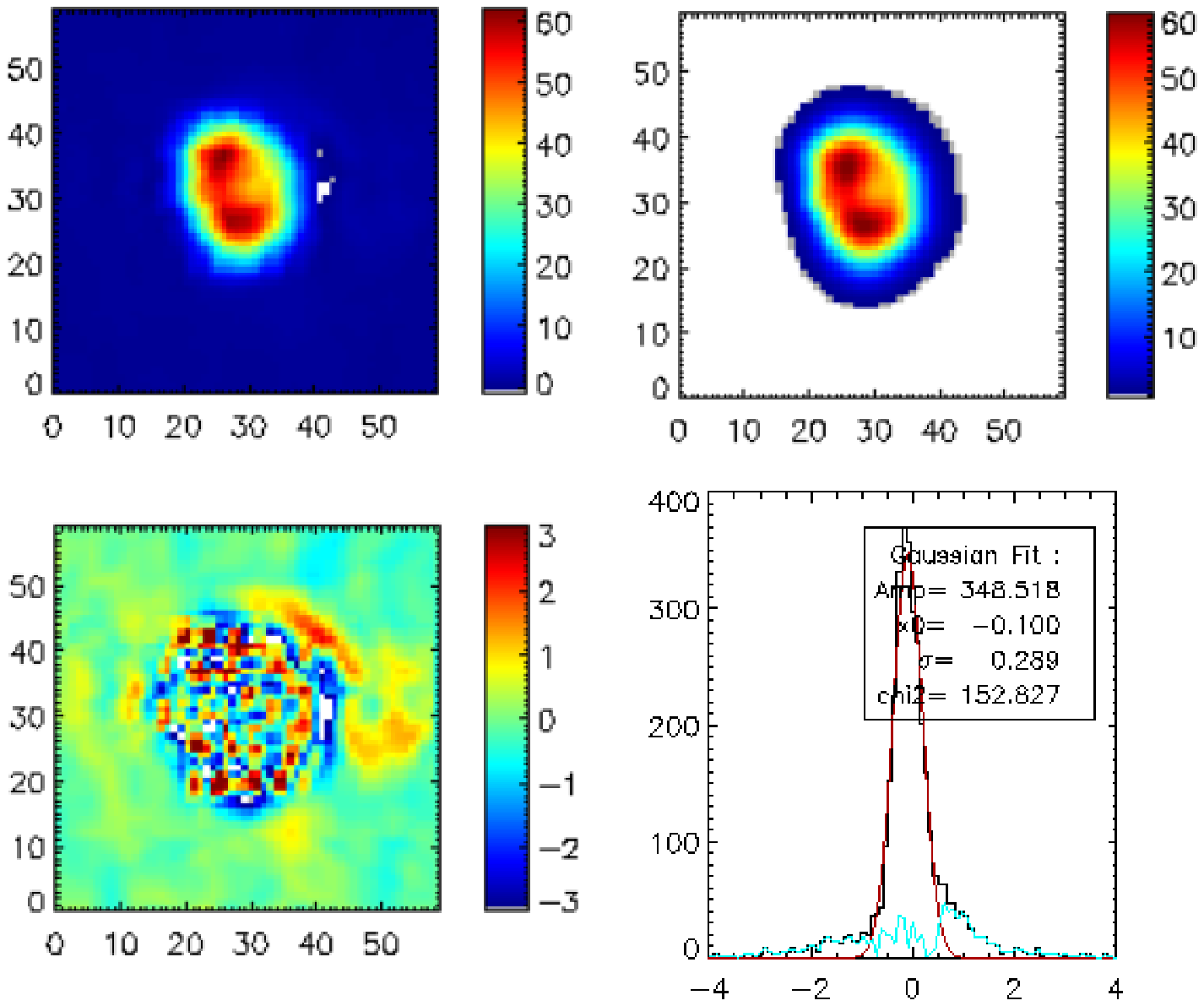}}
  \resizebox{\hsize}{!}{\includegraphics[]{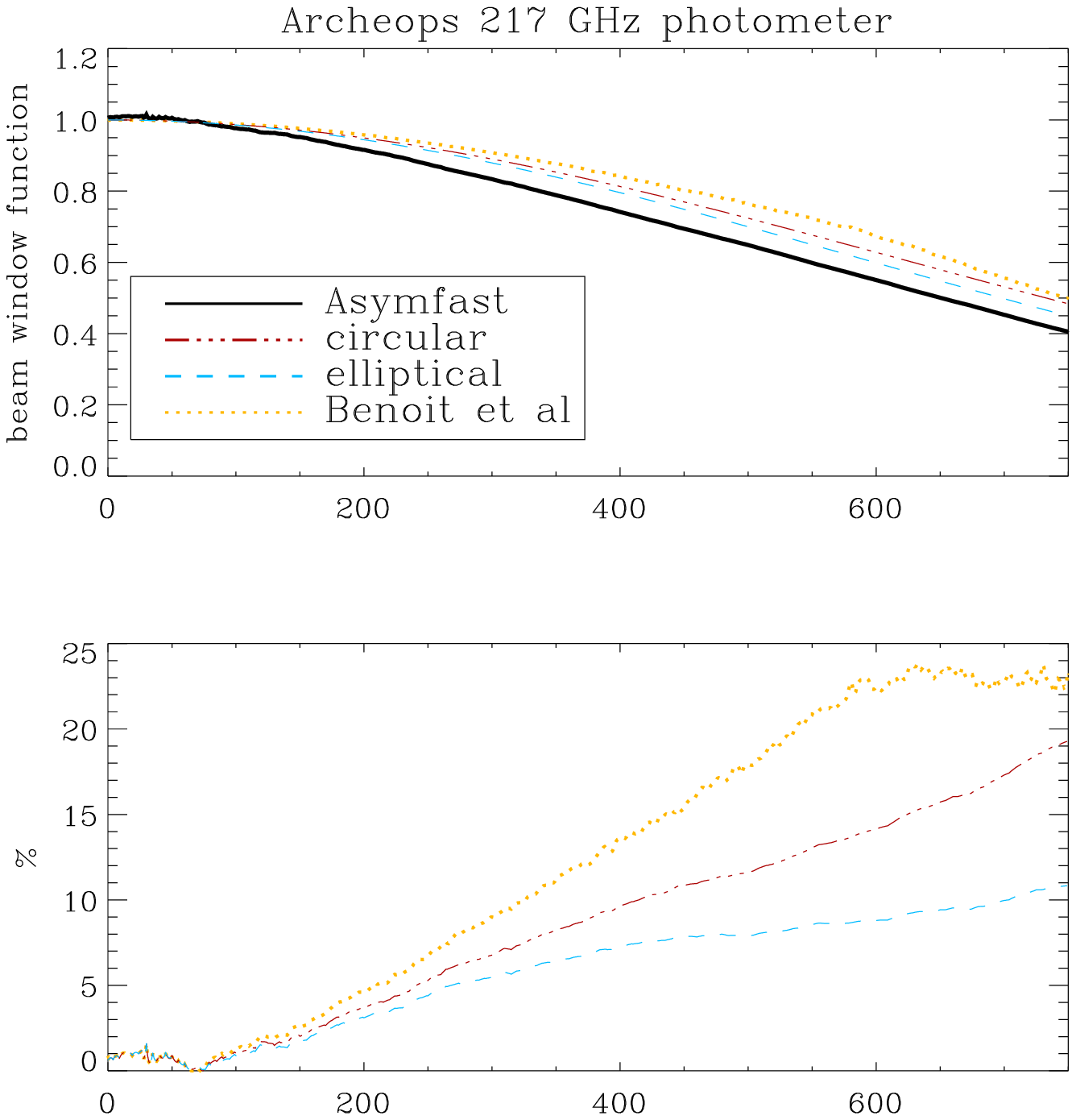}}
  \caption{From left to right and from top to bottom. Main beam map of
    the 217~GHz photometer used in the first Archeops CMB analysis and
    its reconstruction with {\AF} using 10~Gaussians the
    $B_{\ell}^{eff}$; the residuals-map to this one and the residuals
    histogram; the beam transfer function for the Gaussian
    approximation, for the simulations as done in [6] and for {\AF};
    relative differences to the {\AF} beam transfer function
    estimatate for the other three approaches. See text for details.
    \label{archeops}}
\end{figure}

{\AF} has been applied to the main beam of the Archeops bolometers
which are identical to the Planck-HFI ones. The beam shapes were
measured on Jupiter \cite{beamarcheops} and are for most of them
moderately elliptical, the multimode ones being rather
irregular. Figure~\ref{archeops} shows the study of the Archeops
217~GHz photometer used for the first Archeops CMB analysis
\cite{archeops}. On the first row we show the $1 \times 1$~deg with
1~arcminute pixel maps of the main beam, at left the initial map, at
right the reconstructed one with 10~Gaussians. The second row presents
the residual map and its histogram. The latter is fitted with a
Gaussian (in red), the difference to it is shown in blue. The
Monte-Carlo estimation using {\AF} described in Section~\ref{bell}
(solid line) is compared to the analytic transfer function of a
circular approximation of the beam (semi-dashed line), to the
elliptical approximation to order 4 from~\cite{pablo} (dashed line)
and to the estimation obtained by simulation used in \cite{archeops}
(dotted line). They are shown on the third row figure. The bottom
panel shows the differences to the {\AF} beam transfer function
estimation for the other three approaches. 

The beam is well reproduced as the residual map do not exhibit any
structure. Moreover the dispersion of the distribution of the residuals is
compatible to the noise level of the initial map. The deviation to the
Gaussian and to the elliptical models arises from $\ell \approx 200$,
{\it i.e.} the top of the first acoustic peak and culminates around
$\ell \approx 500-600$, {\it i.e.} at the second peak
location. Therefore the fine structure of main beams with equivalent
FWHM of 13~arcminutes is already determinant for the measurement of
the second acoustic peak with Archeops. Because of the irregularity of
the beam, we are more sensitive to higher resolution structures.

\section{Conclusions}
{\AF} is a fast and accurate convolution procedure particularly
well-adapted to asymmetric beam patterns and complex scanning
strategies which are often used in CMB observations. {\AF} can both
produce convolved maps from input timelines and compute, from
Monte-Carlo simulations, an accurate circular approximation to the
transfer function, $B_{\ell}^{eff}$, of any asymmetric beam
pattern. The computing time needed to obtain a convolved map is
dominated by the HEALPix software computing time. So it scales as
\order{n_{pix}^{3/2}} where $n_{pix}$ is the number of pixels of
the map, with a multiplying factor depending on the number of
Gaussians.

{\AF} models any general beam pattern by a linear combination of
circular 2D Gaussians, permitting an accurate reconstruction of the
instrumental beam, with residuals smaller than 1~\% (compared to 4~\%
for an elliptical Gaussian model). In addition, {\AF} convolution is
at least a factor of 50 faster than the brute-force convolution
algorithm for full-sky maps of 12.5 million pixels and even faster at
higher resolution. This allows us to perform a large number of
Monte-Carlo simulations in a reasonable computing time to estimate
accurately the effective circular transfer function of the beam
pattern.

By constrast to other modelling techniques like \cite{page} and
\cite{pablo}, {\AF} can be used with non-circular and non-elliptical
beam patterns. Furthermore {\AF} approximates the main beam pattern while
\cite{wandelt} uses an exact 4-$\pi$ beam description, nevertheless it
can be equally easily applied to any general scanning strategy while the
feasibility of \cite{wandelt} strongly depends on the former \cite{challinor}.

Note that {\AF} is a general convolution algorithm which can be also
used successfully in many other astrophysical areas to reproduce the
effects of asymmetric beam patterns on sky-maps and to compare
observations from independant instruments which requires the
cross-convolution of the datasets. Any circular functions with
analytic description in the harmonic space may be used instead of
2D~symmetric Gaussians.

%________________________________________________________________
\begin{acknowledgments}
  The authors would like to thank F.-X.~D\'esert for fruitfull
  discussions. The Healpix package was used throughout the data
  analysis~\cite{healpix}. We also want to thank the anonymous referee
  for pointing out some issues which helped us to improve the manuscript.
\end{acknowledgments}
%________________________________________________________________

%________________________________________________________________

\end{document}